# Optimizing the Whole-life Cost in End-to-end CNN Acceleration

Jiaqi Zhang, Xiangru Chen, and Sandip Ray

University of Florida

*Abstract*—The acceleration of CNNs has gained increasing attention since their success in computer vision. With the heterogeneous functional layers that cannot be processed by the accelerators proposed for convolution layers only, modern end-to-end CNN acceleration solutions either transform the diverse computation into matrix/vector arithmetic, which loses data reuse opportunities in convolution, or introduce dedicated functional unit to each kind of layer, which results in underutilization and high update expense.

To enhance the whole-life cost efficiency, we need an acceleration solution that is efficient in processing CNN layers and has the generality to apply to all kinds of existing and emerging layers. To this end, we propose GCONV Chain, a method to convert the entire CNN computation into a chain of standard general convolutions (GCONV) that can be efficiently processed by the existing CNN accelerators. This paper comprehensively analyzes the GCONV Chain model and proposes a full-stack implementation to support GCONV Chain. On one hand, the results on seven various CNNs demonstrate that GCONV Chain improves the performance and energy efficiency of existing CNN accelerators by an average of 3.4x and 3.2x respectively. On the other hand, we show that GCONV Chain provides low whole-life costs for CNN acceleration, including both developer efforts and total cost of ownership for the users.

## 1. Introduction

Since its resurgence, Convolutional Neural Network (CNN) has demonstrated impressive success in promoting the computer vision in a wide range of applications [1]–[4]. However, the high accuracy of CNNs is achieved at the cost of enormous computation and data movement, which is an undesirable obstacle to widely implementing and deploying them. Therefore, CNN acceleration has received increasing attentions.

Normally, CNN computation and parameters are dominated by the convolution layers. Based on this fact, abundant prior works [5], [6], [15]–[17], [7]–[14] focus on the acceleration of these layers by designing customized architectures and dataflows to enhance the performance and data reuse in convolution operations (we classify these accelerators as **CIP**, i.e., convolution intended processors). However, recent CNNs incorporate more heterogeneous functional layers. For example, Figure 1(a) depicts a basic block of MobileNet [18] with four various layers. Except for the first layer, each of them performs unique computation that cannot fit into the traditional definition of a convolution layer and thus cannot be accelerated by CIPs as illustrated in Figure 1(b). Since these non-traditional layers play an important role in promoting the accuracy [19] of CNNs and are even proved to have better learning capability than the traditional layers [20], overlooking them can lead to degraded accuracy. Therefore, CIPs that only accelerate the convolution layers and are even incapable of parsing the parameters of other layers have to offload them to somewhere else, failing to efficiently perform the end-to-end CNN computation.

To deal with the "elephant in the room", a common practice is to transform the computation in diverse functional layers into tensor (mostly matrix and vector) operations and perform them in the tensor instruction processors (**TIP**) [21]–[24]. As illustrated in Figure 1(c), TIPs are able to process any CNN layer but they lose certain data reuse opportunities compared with CIPs and thus result in low data movement efficiency. TIPs also suffer from lower code density because they explicitly instruct the data loading and each matrix/vector operation. Another plausible solution is to add a dedicated on-chip processing unit for each type of layer in addition to the convolution (these accelerators are categorized as **LIP**, i.e., the layer instruction processors) [25]–[27], as illustrated in Figure 1(d). Nevertheless, almost every new network features a new technical layer. It is costly for the hardware developers to design a new component to process the newly introduced layer for the full-fledged accelerators and for the users to upgrade their deployment. What's more, the varying CNN structures make it impossible to design a heterogeneous accelerator with high utilization for all the workloads.

To help understanding the above-mentioned accelerators, we analogize TIPs as RISC and compare CIPs and LIPs to CISC. The challenges faced by the existing works drive us to find *a balance point between the generality of RISC-like and efficiency of CISC-like CNN accelerators*. We innovatively propose to convert the end-to-end diverse CNN computation into a series of general convolution (**GCONV Chain**), where the operations are as uniform as that in TIPs and the reuse opportunities can be leveraged as in CIPs and LIPs, as shown in Figure 1(e). To be more specific, though the diverse layers perform different functions, we develop a parameterized model that expresses all the CNN operations as a scale-up of a simple 1-D GCONV. For example, the local response normalization [28] and batch normalization [19] can be viewed as general convolutions in the channel dimension and batch dimension respectively. This enables the end-to-end CNN computation to be efficiently accelerated by a uniform computation fabric, where the complex acceleration problem is simplified to a single mapping algorithm of the 1-D GCONV.

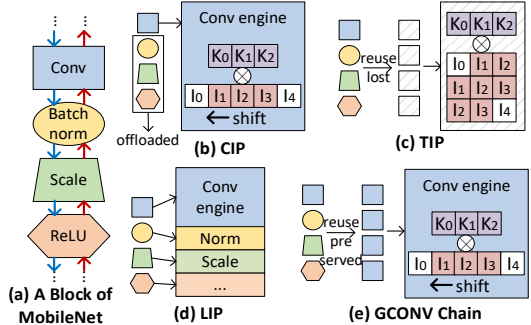

Figure 1. Three Types of Modern CNN Accelerators and GCONV Chain

Table 1. Impact of Non-Traditional Layers on Modern CNN Acceleration

| (a) Non-Traditional Layers in CNNs | | | | | | | (b) Inefficiencies of Accelerators | | |
|---|---|---|---|---|---|---|---|---|---|
| CNN type | Net-work | New layer types | Non-traditional ratio | | | | Data repli-ca-tion (TIP) | Of-float-ing (CIP) | Utili-za-tion (LIP) |
| | | | Lay-ers | Com-puta-tion | Data foot-print | Move-ment | | | |
| clas-sifi-ca-tion | AN | LRN, dropout | 24% | <1% | 5% | <1% | 35x | 3% | 98% |
| | GLN | ave pool, concat | 13% | <1% | 17% | 1% | 6x | 13% | 80% |
| | DN | batch norm, scale | 66% | 5% | 76% | 22% | 2x | 53% | 17% |
| | MN | depthwise conv | 62% | 8% | 73% | 16% | 2x | 77% | 11% |
| other | ZFFR | RoI, proposal | 29% | <1% | 41% | <1% | 4x | 57% | 86% |
| | C3D | 3D conv, 3D pool | 52% | 99% | 46% | 99% | 6x | 87% | 1% |
| | CapNN | prim, digicaps | 18% | 95% | 6% | 93% | 3x | 33% | 1% |

The benefits of our work are threefold as will be evaluated in Section 6: (1) since GCONV Chain is composed of general convolutions and CNN accelerators are designed to perform convolution, it can be executed by almost any CNN accelerator with no worse if not better performance (Section 6.3); (2) when GCONV Chain is applied to CIPs, it eliminates the costly offloading overhead. And this GCONV armed CIP (GC-CIP) does not suffer from the low data movement efficiency in TIPs and resource underutilization in LIPs, showing that GC-CIPs are promising accelerators for the entire end-to-end diverse CNN computation (Section 6.5); (3) the generality of GCONV Chain further makes GC-CIPs the most whole-life cost efficient choice by lessening the burden of both users and developers (Section 6.6).

In summary, this paper makes the following contributions:

(1) We recognize the gap between modern CNNs composed of diverse functional layers and the existing CNN accelerators which are proposed for either generality or efficiency for certain layers.

(2) We propose GCONV Chain, a method to convert the end-to-end CNN computation into a chain of general convolutions (GCONVs) so that all the CNN layers can be efficiently and uniformly accelerated.

(3) We further propose a full-stack implementation of GCONV Chain on existing CNN accelerators, including GCONV Chain generation, acceleration and hardware support.

(4) Our experiments on seven popular CNNs from a wide range of applications and five representative accelerators of different types show that GCONV Chain can significantly improve the performance and energy efficiency of existing CNN accelerators and benefit the cost efficiency of both development and ownership.

The rest of this paper is organized as follows. Section 2 provides the background on CNN acceleration and the motivation for CNN generalization. Section 3 describes the approach to transform CNN computation into a GCONV Chain. Then Section 4 introduces the methodologies to analyze the GCONV workload and the algorithm to accelerate GCONV Chain. Section 5 elaborates the system implementation of GCONV Chain. Sections 6 evaluates our proposed GCONV Chain. And Sections 7 and 8 discussed the related work and concludes the paper respectively.

## 2. Background and Motivation
### 2.1 A Traditional Convolution Layer

CNNs are neural networks featured by the dominating convolution layers. Figures 2(a) and (b) provide the illustration and pseudo code of a traditional convolution layer to help understand the computation pattern and reuse opportunities in the convolutional operation, where $Np$ stands for the argument of parameter $p$. The same set of $Noc$ kernels are applied to a batch of $Nbs$ input samples. The size of each kernel is $Nky \times Nkx \times Nic$ and they slide on the height and width of the input channels to generate $Noc$ channels of $Noy \times Nox$ outputs. In some layers of certain CNNs [8][9], the input channels are divided into $Ngp$ groups to perform convolution individually.

There are two kinds of reuses in convolution, *parallel-reuse* and *overlap-reuse*. Parallel-reuse results from parallel computation in terms of a certain data. For example, each kernel is reused to generate outputs within each channel and the inputs are reused by different output channels. Similarly, each output is reused within the kernel since the partial results are reduced to generate one output. The sliding of the kernel also brings overlap-reuse when the input windows of consecutive outputs are overlapped. For instance, in Figure 1(a), the inputs with dots are shared by the computation of different outputs, multiplied by different weights. Exploiting these two kinds of parallelisms in the convolutional operation has been widely recognized as an effective method to reduce the costly data movement in CNN acceleration [29].

### 2.2 Non-Traditional Layers in CNNs

Aside from the convolution layers, there is a variety of other necessary layers in the CNNs, each of which performs a different operation on the input data. And the operation of a same layer even varies in the training and inference of the network. Besides the traditional layers such as fully connection, max pooling, ReLU and softmax that breed the success of the nascent CNN model, i.e., LeNet [30], the growing power of the emerging CNNs is always accompanied with new functional layers. Benchmarked on seven popular CNNs in the timeline (i.e., four classification CNNs: AlexNet (AN) [28], GoogLeNet (GLN) [31], DenseNet (DN) [32], MobileNet (MN) [18], in addition to an object detection network Faster R-CNN+ZFNet (ZFFR) [33][34], a 3-D CNN for video processing (C3D) [35] and a capsule neural network (CapNN) [36]), Table 1(a) lists the newly introduced layers and the ratio of these non-traditional layers. As can be observed in Columns 5 to 7, the non-traditional layers account for increasing computation, data footprint and movement in the CNNs. And these layers are taking significant roles in determining the speedup and energy efficiency of CNNs.

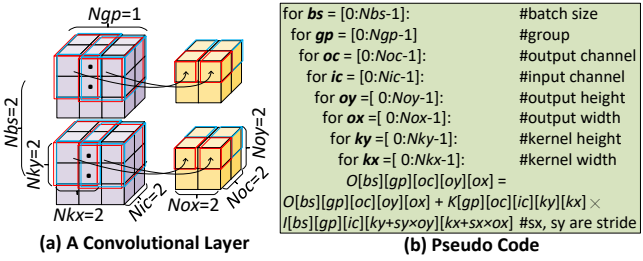

Figure 2. A Traditional Convolution Layer

## 2.3 Modern CNN Acceleration Challenges

As CNNs are being employed more frequently and the heterogeneity in CNN layers keeps increasing, it is obviously important to have a solution that efficiently accelerates all the CNN computation. Unfortunately, we notice that almost all the existing works suffer from various inefficiencies when processing modern CNNs.

**TIP:** The tensor instruction processors (TIP) are able to process any CNN layer by transforming them into tensor (mostly matrix and vector) operations [21]–[24]. However, they cannot exploit the abundant overlap-reuses in CNNs. For instance, *im2col* [22] is commonly employed by TIPs to transform convolution into matrix multiplication, where the input windows are flattened to columns in a matrix and then multiplied by a weight matrix, as shown in Figure 1(c). This results in the replication of the input data (marked in red) and thus low energy efficiency. Column 1 in Table 1(b) quantifies the total data replication of the CNNs in TIPs. Since the power consumption is dominated by the data movement [24], this overhead significantly increases the operating expense for the users.

**CIP:** The main component of a convolution intended processor (CIP) is a convolution engine, which implements various exhaustively explored dataflows (e.g., weight stationary, output stationary, row stationary, etc.) to maximally exploits both the parallel and overlap-reuse opportunities in convolution [5], [6], [15]–[17], [7]–[14]. For example, the convolution engine in Figure 1(b) adopts the weight stationary dataflow, where the inputs are shifted along the stationary weights to exploit the overlap-reuse, avoiding data replication. However, since the proposed dataflows only apply to the computation model of traditional convolution layers, CIPs are inefficient or even incapable of parsing the parameters when processing the other layers. Therefore, offloading is required for non-traditional layers whose acceleration is omitted in CIPs. Column 2 in Table 1(b) characterizes the ratio of intermediate data that requires offloading for a series of non-traditional processing. Note that since the offloading energy consumption can be as high as 146x of the on-chip data movement in our experiment, this adds considerable burden to the system.

**LIP:** The layer instruction processors (LIP) [25]–[27] add a dedicated on-chip processing unit for each type of layer in addition to the convolution engine and process the corresponding layers of different inputs in a pipeline. Resulted from the variation of the number and shape of each kind of layers in a certain CNN, pipeline bubbles are unavoidable. To implicate this, Column 3 in Table 1(b) lists the utilization of different networks assuming the pipeline has two stages for traditional and non-traditional layers respectively. The resources are partitioned based on the ratio of the traditional and non-traditional computation in all the networks. As observed, the uniform partitioning results in significantly varying utilization. And the utilization is extremely low in networks with more non-traditional computation (e.g., C3D and CapsuleNN). Layers that pose a pipeline barrier also lower the utilization (e.g., batch normalization in DenseNet and MobileNet). What's more, LIPs demand the developers to design an efficient acceleration solution to any new layer and the users to update the devices as frequently as the evolution of CNNs.

To enhance the whole-life cost efficiency of end-to-end CNN computation, we need an acceleration solution that is efficient in processing CNN layers and has the generality to apply to all kinds of existing and emerging layers.

## 3. General Convolution (GCONV) Chain

Our goal is to convert the diverse CNN layers into a chain of standard operations without losing the convolution pattern. Therefore, instead of breaking them down into basic matrix/vector arithmetic as in TIPs, we view them as multi-dimension convolutions under a more generalized definition. This section proposes the GCONV model and the method to transform the entire CNN computation into a GCONV Chain.

### 3.1 GCONV Operation Definition

GCONV is the most basic operation in our system. It is a concisely parameterized 1-D convolution which can be scaled up to multiple dimensions to define various CNN operations. Compared with the traditional definition of a convolution layers, the *simplicity*, *scalability* and *representability* of GCONV make it ideal to effortlessly, uniformly and efficiently accelerate all kinds of CNNs.

**Definition:** Figure 3 shows an example of 1-D GCONV and the definition of the parameters. It is characterized by four parameters that form the nested loop of the computation as shown in Figure 4: the inputs are separated into $Ng$ groups and no connection or reuse exists inter-group; in each group, the inputs are convolved by $Nop$ kernels in parallel; each kernel has $Nks$ weights; and $Nopc$ outputs are generated by each kernel. There are also two auxiliary parameters, i.e., *ps* for padding size and *s* for stride, as in the traditional convolution. The input size $Nipc$ is not required since it can be derived as:

$$Nips = (Nopc + 1) \times s + Nks - 2ps \quad (1).$$

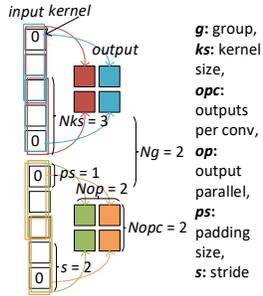
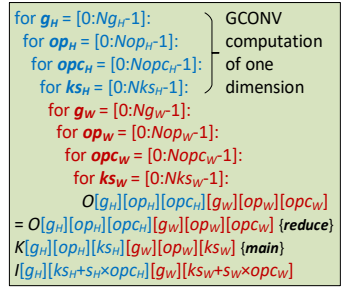
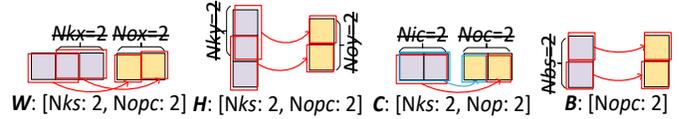

Figure 3. 1-D GCONV

Figure 4. Pseudo Code of 2-D GCONV in H and W Dimensions

Figure 5. GCONV Definition of the Convolution Layer in Figure 2

**Simplicity:** Recall that the traditional convolution layer in Figure 2 has eight different parameters. The 1-D GCONV is relatively simple with only four. And these four parameters are necessary to preserve the data reuse patterns discussed in Section 2.1. Specifically, the input, kernel and output can be parallel-reused for *Nop*, *Nopc*, *Nks* times and the input can be overlap-reused by the outputs when $Nks > s$. In our system, if not explicitly denoted, the parameters take their default values as [*ps*: 0, *s*: 1, *Ng*: 1, *Nop*: 1, *Nks*: 1, *Nopc*: 1].

**Scalability:** This simple 1-D GCONV can be easily scaled up to multi-dimension GCONVs that are actually processed in the accelerators. The dimension of GCONV is determined by the dimension of the data in the network. For example, the convolution layer in Figure 2 manifests four dimensions: the mini-batch (***B***), channel (***C***), height (***H***) and width (***W***). The operation can be described as a 1-D GCONV in each dimension, as shown in Figure 5, where the traditional convolution parameters are replaced by the GCONV parameters. While the GCONV operations in dimensions *H* and *W* are natural, explanations are needed to understand the 1-D GCONVs in the other two dimensions, which might be counter-intuitive at first. Specifically, in dimension *C*, the kernels cover the entire inputs in parallel, which means the operation of each kernel can be viewed as a convolution with kernel size equal to the input size. As for dimension *B*, the same kernel is applied to all the samples in a mini-batch. This can be viewed as a kernel with one weight moving along the inputs by a stride of 1.

One benefit of replacing the 4-D integrated operation with the scaled-up 1-D GCONV is the outstanding scalability. Figure 4 lists the nested loop for multi-dimension GCONV. Note that the exact same four loops are duplicated (marked as red and blue) for each new dimension, so it is easily inferred that the 4-D GCONV is a nest of 16 loops. And we can remove the four loops in dimension *B* to model the real-time learning [37], duplicate four loops for time dimension in 3-D CNNs [38] or for vector dimension in Capsule Neural Networks [36], etc. Although the number of loops in theory seem huge, it can be pruned if a certain parameter takes its default value. Therefore, the GCONV model does not bring any overhead to the actual number of parameters or effectual loops, as illustrated in Figure 5. On the contrary, since all the dimensions are perfectly symmetric in terms of the computation and data reuse opportunities, it indeed shrinks our analysis and acceleration design space to just 1-D GCONV with only four loops.

**Representability:** Besides, the proposed model can further shrink the effort-taking study of all kinds of layers in both training and inference into a 1-D GCONV. In a comprehensive analysis, we find that all the layers in modern CNNs can be decomposed into a series of GCONV operations (an example will be given in Section 3.2). And it is future-proof because GCONV can always model a tensor operation by setting the kernel size equal to the input size.

We also notice that although the computation pattern of all the layers can be represented by the GCONV parameters, not all of them perform *multiply*-and-*add* operations. Therefore, GCONV operators are introduced aside from the parameters. The four operators of respectively define how the inputs are preprocessed (***pre***); how they are processed by the kernels (***main***); how the partial results within a kernel are reduced (***reduce***); and how the outputs are postprocessed (***post***). The *pre* or *post* operator applies the same processing (e.g. *multiply, and, square* or *LUT*) to each input and output when they are loaded into the convolution engine or sent back to the global buffer. The convolution engine convolves the inputs with the kernels and performs *main* operation (could be *square, multiply, and* or *add*) between inputs and kernel parameters (they are no longer named "weights" because the operation is not limited to *multiply*). Some connections in the convolution engine allow the *reduction* (could be *add* or *compare*) of partial results spatially or temporally. The operators are the same across all the dimensions in a GCONV operation.

Since GCONV does not modify the dataflow, almost all the existing CNN accelerators can support GCONV computation. The only modification required is that the accelerators, which only perform *multiply* and *add* in traditional convolution, should be equipped with more comprehensive functions for *main* and *reduce* operators. This only brings little overhead compared to the expensive dataflow implementation, as will be evaluated in Section 6.4. Note that we focus on the computation pattern of the CNN layers, so the optimization of the single operators is not within the scope of this paper.

Table 2. GCONVs for Batch Normalization Layer (*Nbs*: mini-batch size, *Nic*: number of input channels, *Noy*/*Nox*: number of outputs in the *H*/*W* dimension per channel, *I*/*O*: input/output of the layer, $gI$/$gO$: gradient of *I*/*O*, L(*l*-1)/(*l*+1): the last/next layer)

| GCO NV | GCONV Parameters | | | | Input | Kernel Param | Operators | | | | Computation |
|---|---|---|---|---|---|---|---|---|---|---|---|
| | *B* | *C* | *H* | *W* | | | *pre* | *main* | *reduce* | *post* | |
| **FP** | | | | | | | | | | | |
| **FP1** | [*Nks*: *Nbs*] | [*Nopc*: *Nic*] | [*Nopc*: *Nix*] | [*Nopc*: *Niy*] | L(*l*-1)_output | | | | + | ×1/*Nbs* | $\mu = \sum I/Nbs$ |
| **FP2** | [*Nopc*: *Nbs*] | [*Ng*: *Nic*] | [*Ng*: *Nix*] | [*Ng*: *Niy*] | L(*l*-1)_output | FP1_output | | − | | | $t1 = I - \mu$ |
| **FP3** | [*Nks*: *Nbs*] | [*Nopc*: *Nic*] | [*Nopc*: *Nix*] | [*Nopc*: *Niy*] | FP2_output | | ^2 | | + | LUT | $t2 = 1/\sqrt{\sum t1^2/Nbs + \varepsilon}$ |
| **FP4** | [*Nopc*: *Nbs*] | [*Ng*: *Nic*] | [*Ng*: *Nix*] | [*Ng*: *Niy*] | FP2_output | FP3_output | | × | | | $O = t1 \times t2$ |
| **BP** | | | | | | | | | | | |
| **BP1** | [*Nks*: *Nbs*] | [*Nopc*: *Nic*] | [*Nopc*: *Nix*] | [*Nopc*: *Niy*] | L(*l*+1)_gradient | FP4_output | × | | + | ×1/*Nbs* | $t3 = \sum O \times gO/Nbs$ |
| **BP2** | [*Nopc*: *Nbs*] | [*Ng*: *Nic*] | [*Ng*: *Nix*] | [*Ng*: *Niy*] | BP1_output | FP4_output | | × | | | $t4 = O \times t3$ |
| **BP3** | [*Nks*: *Nbs*] | [*Nopc*: *Nic*] | [*Nopc*: *Nix*] | [*Nopc*: *Niy*] | L(*l*+1)_gradient | | | | + | ×1/*Nbs* | $t5 = \sum gO/Nbs$ |
| **BP4** | [*Nopc*: *Nbs*] | [*Ng*: *Nic*] | [*Ng*: *Nix*] | [*Ng*: *Niy*] | L(*l*+1)_gradient | BP3_output | | − | | | $t6 = gO - t5$ |
| **BP5** | [*Ng*: *Nbs*] | [*Nopc*: *Nic*] | [*Nopc*: *Nix*] | [*Nopc*: *Niy*] | BP4_output | BP2_output | | − | | | $t7 = t6 - t4$ |
| **BP6** | [*Nopc*: *Nbs*] | [*Ng*: *Nic*] | [*Ng*: *Nix*] | [*Ng*: *Niy*] | BP5_output | FP3_output | | × | | | $gI = t7 \times t2$ |

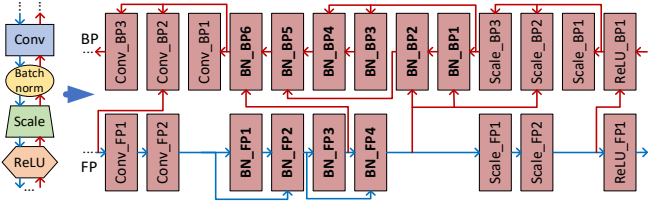

Figure 6. GCONV Chain of the MobileNet Block in Figure 1(a)

### 3.2 GCONV Chain Formation

With a stack of different functional layers, the end-to-end CNN computation can be converted into a GCONV Chain based on the producer and consumer relations.

First, here is an example of how to transform the batch normalization (BN) layer in both forward (FP) and back (BP) propagation into a GCONV Chain. In FP, the outputs (*O*) are the normalization of the inputs (*I*) over the entire mini-batch:

$$O_i^b = \frac{I_i^b - \mu_i}{\sqrt{\sigma^2_i + \varepsilon}} \quad (2),$$

where *b* denotes the index in *B* dimension, *i* iterates in the *C*, *H*, *W* dimensions and *ε* is a small parameter. $\mu_i$ and $\sigma^2_i$ are the mean and variance of the mini-batch (the size is *Nbs*):

$$\mu_i = \frac{\sum_{b=0}^{Nbs-1} I_i^b}{Nbs} \quad (3),$$

$$\sigma^2_i = \frac{\sum_{b=0}^{Nbs-1} (I_i^b - \mu_i)^2}{Nbs} \quad (4).$$

Table 2 and the GCONVs in bold in Figure 6 show the GCONV Chain of BN. The FP GCONVs are generated by analyzing the dependencies of Equations (2) to (4). Since both *O* and $\sigma^2$ depend on $\mu$, the calculation of $\mu$ is first appended to the chain (**FP1**), which is a reduction in the *B* dimension. Then $(I - \mu)$ is calculated next **(FP2)** as a GCONV with no reduction but different kernel parameters for each data in *C*, *H*, *W* dimensions. After that, the calculations of $\sigma^2$ and O are appended as **FP3** and **FP4** sequentially.

The BP of BN performs the following operation:

$$gI_i^b = \sum_{bb=0}^{bs-1} \frac{\partial O_i^{bb}}{\partial I_i^b} gO_i^{bb}$$

$$= (gO_i^b - \underbrace{\underbrace{\sum_{bb=0}^{bs-1} \frac{gO_i^{bb}}{bs}}_{BP3} - O_i^n \underbrace{\sum_{bb=0}^{bs-1} \frac{O_i^{bb} gO_i^{bb}}{bs}}_{BP1})/\sqrt{\sigma^2_i + \varepsilon}}_{\underbrace{\underbrace{\phantom{xx}}_{BP4} \underbrace{\phantom{xx}}_{BP2}}_{\underbrace{BP5}_{BP6}}} \quad (5),$$

which is similarly decomposed into six GCONVs. The GCONVs of other layers can also be derived in this way.

Then based on the inter-layer dependencies, we can establish the chain for the entire CNN. Figure 6 illustrates the conversion of the block in Figure 1(a) to a GCONV Chain. Like CISC instructions, the original block contains a pile of diverse layers, each requiring complicated customized analysis and optimization. Cleverly, our proposed technique works as a "*micro code*" layer that translates them into a chain composed of only GCONV operations. It might be noticed that the code density increases in Figure 6. To this end, we will introduce an operation fusion technique in Section 4.3.

## 4. GCONV Chain Acceleration

As shown in Figure 6, the generalization approach proposed in Sections 3 allows diverse operations in CNNs to be converted into a chain of standard GCONV operations. This way, the acceleration is no longer layer-specific. Instead, the entire CNN acceleration can be uniformly and systematically achieved by studying the acceleration of a single GCONV operation (Section 4.1) and optimizing the interaction between operations on the chain (Section 4.3).

### 4.1 Mapping a Single GCONV Operation

To accelerate the GCONV, we need to perform more computation in parallel while increasing the data reuse, which is realized by unrolling and exchanging the order of the nested loop in Figure 4. The loops can be unrolled both spatially and temporally in an accelerator. Different accelerators have different spatial dimensions to unroll the computation and different memory structures for temporary data storage. Here we build an example on Eyeriss [5], one of the most complicated and self-contained CNN accelerators. The generality of the GCONV mapping will be discussed in Section 4.4.

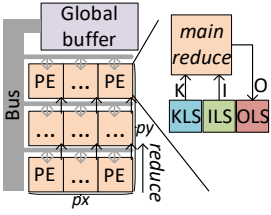
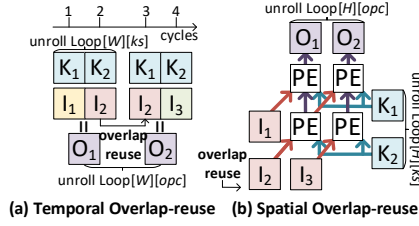
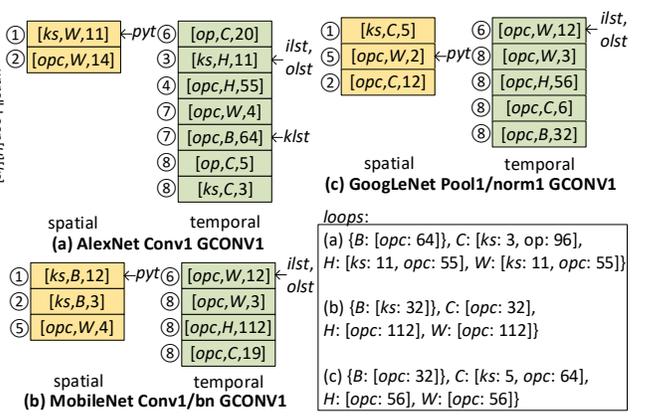

Figure 7. Eyeriss Structure    Figure 8. Eyeriss Overlap-reuse Primitives

**Algorithm 1**: Algorithm for GCONV Mapping on Eyeriss
**Input:** GCONV loops of four parameters in four dimensions *loops*; accelerator PE array size $py = 12$, $px = 14$; LS size $ils = 12$, $kls = 224$, $ols = 24$ [5].
**Output:** two unrolling lists *spatial* and *temporal*.
1: **function** unrolling (*ud*, *p*, *d*)
2:    $uf \leftarrow$ **min** (remaining resources, *loops*[*d*][*p*])
3:    *loops*[*d*][*p*] $\leftarrow$ **ceil** (*loops*[*d*][*p*]/*uf*)
4:    remaining resources $\leftarrow$ **floor** (remaining resources/*uf*)
5:    **return** *uf*
6: **function** main ()
7:    **for** *d* **in** ["*W*","*H*","*C*","*B*"] **do**
8:      **if** overlap-reuse **then**
9:        *spatial*.**append** (["*ks*", *d*, unrolling("*py*", "*ks*", *d*)])
10:        *spatial*.**append** (["*opc*", *d*, unrolling("*px*", "*opc*", *d*)])
11:      **if** second overlap-reuse **then**
12:        *temporal*.**append** (["*ks*", *d*, unrolling("LS","*ks*", *d*)])
13:        *temporal*.**append** (["*opc*", *d*, *loops*[*d*]["*opc*"]])
14:    **for** *p* **in** ["*ks*","*opc*","*op*","*g*"] **do**
15:      **for** *d* **in** ["*W*","*H*","*C*","*B*"] **do**
16:        *spatial*.**insert** ([*p*, *d*, unrolling("*py*", *p*, *d*)])
17:    **for** *p* **in** ["*opc*","*op*","*ks*","*g*"] **do**
18:      **for** *d* **in** ["*W*","*H*","*C*","*B*"] **do**
19:        *spatial*.**append** ([*p*, *d*, unrolling("*px*", *p*, *d*)])
20:    **for** *p* **in** ["*op*","*ks*", "*opc*", "*g*"] **do**
21:      **for** *d* **in** ["*W*","*H*","*C*","*B*"] **do**
22:        *temporal*.**insert** ([*p*, *d*, unrolling("LS", *p*, *d*)])
23:    **for** *p* **in** ["*opc*","*op*", "*ks*", "*g*"] **do**
24:      **for** *d* **in** ["*W*","*H*","*C*","*B*"] **do**
25:        *temporal*.**append** ([*p*, *d*, *loops*[*d*][*p*]])

*ud*: accelerator unrolling dimension, *uf*: unrolling factor, LS: local scratchpads.
If *uf* is 1, do not append or insert.

Figure 9. Example Unrolling Lists (Mini-batch Size is 32)

**Accelerator structure:** Figure 7 shows the on-chip structure of Eyeriss. For neatness, we focus on the abstracted spatial and temporal unrolling dimensions while omitting the other implementation details that do not affect the GCONV mapping. It contains a $py \times px$ PE array and a global buffer which communicates with the PEs via broadcasting. Each PE consists of a *main* and *reduce* (multiply and add in the original work) unit in addition to three local scratchpads for inputs (ILS), kernel parameters (KLS) and outputs (OLS) to reduce global buffer access.

First, the loops can be spatially unrolled vertically (*py*) or horizontally (*px*) in the PE array. The spatial unrolling determines the parallelization of the computation and the spatial data reuse. The input and kernel parameter parallel-reuses are enabled both horizontally and vertically. The partial results can only be *reduce*d (i.e. output parallel-reuse) vertically thanks to the forwarding links between the rows. Second, the loops can be unrolled temporally so that each PE can reuse the data or *reduce* the partial results locally in LS.

Like many accelerators proposed for convolution layers, the original work of Eyeriss provides overlap-reuse primitives for *W* and *H* dimensions (i.e., row-stationary). As shown in Figure 8(a), Loop[*W*][*ks*] is unrolled temporally followed by Loop[*W*][*opc*]. This enables the local scratchpads to load only *s* instead of *ks* new inputs each time. In addition, Loop[*H*][*ks*] and Loop[*H*][*opc*] are unrolled in *py* and *px* respectively (Figure 8(b)). This way, the inputs can be shared diagonally in the PE array. In GCONV, these specially-designed primitives will be allocated to any dimension with overlap-reuse instead of being dedicated to *W* or *H*.

**Mapping algorithm:** The algorithm for GCONV mapping in Eyeriss is listed in Algorithm 1. The **main** function is a procedure to append unrolling entries to two unrolling lists, i.e., spatial and temporal, until all the loops are unrolled. Each entry in the lists is [*p*, *d*, *uf*], indicating the unrolling factor of parameter *p* in dimension *d* (Loop[*d*][*p*]). The **unrolling** function determines the unrolling factor of an entry by considering the remaining iterations of the loop and the related PE or LS resources (Lines 2 to 4). Here, we explain Algorithm 1 with example mapping results of three different types of layers in Figure 9, i.e., (a) convolution, (b) batch normalization, (c) local response normalization. Since there are two spatial dimensions, a pointer (*pyt*) is used to point to the tail of the unrolling entries in *py*. Similarly in the temporal unrolling list, three pointers, *ilst*, *olst* and *klst*, point to the last temporal unrolling entries that enable data reuse in ILS, OLS and KLS respectively.

First, to avoid the waste of overlap-reuse primitives, we search for dimensions with overlap-reuse opportunities and unroll *ks* and *opc* in these dimensions in the overlap-reuse primitives (Lines 7 to 13 in Algorithm 1). Note the spatial list is filled before temporal list to maximize parallelism (①②, ③④ in Figure 9). When performing spatial unrolling, the resources in Lines 2 and 4 are simply the PEs. For temporal unrolling, the entailed LS resources are determined by the amount of data of the unrolled tile, which will be discussed in Section 4.2.

Table 3. Data Movement for GCONV

| Data Type | Reuse | Data Movement |
|---|---|---|
| input | $\prod Pop_d$ | $\prod(Pg_d \times (Pks_d + Ps_d \times (Popc_d - 1)))$ |
| kernel param | $\prod Popc_d$ | $\prod(Pg_d \times Pop_d \times Pks_d)$ |
| output | $\prod Pks_d$ | $\prod(Pg_d \times Pop_d \times Popc_d)$ |

After the overlap-reuse primitives, we further fill the spatial unrolling dimensions (Lines 14 to 19) if there are still spare PEs (⑤). It is important to allow the loops that need a certain function to fill the unrolling dimension with that function first. In Eyeriss, *ks* is first unrolled in *py* to exploit the *reduce* function and *opc* and *op* are first unrolled in *px* to exploit the output bandwidth.

Then the loops are unrolled temporally to fill the local scratchpads to increase data reuses (Lines 20 to 22, unrolling entries ⑥). Here, *op* is first unrolled to reuse the inputs. When a local scratchpad (e.g., *kls*) is full, the loops that reuse this kind of data can still be appended (e.g., ⑦).

When all the resources are exploited, the remaining loops are simply appended (Lines 23 to 25, unrolling entries ⑧). Note that *g* is always unrolled the last because it never manifests any special function or data reuse.

### 4.2 Modeling the Performance of GCONV Mapping

To help choosing and evaluating the mapping strategies, this section builds a concise model on how the GCONV mapping results affect the performance and total data movement.

**Computation cycles:** The total cycles to complete a GCONV can be derived from the spatial unrolling as:

$Cyc. = \prod_{d \in \{B,C,H,W\}} \prod_{p \in \{ks,opc,op,g\}} ceil(\frac{Np_d}{SP\_Pp_d})$ (6),

where $Pp_d$ refers to the unrolling factor of parameter *p* in dimension *d* and *SP* means the unrolling in spatial list.

**Data movement:** The total amount of data for a series of unrollings is related to the data reuse opportunities discussed in Section 3.1. As listed in Table 3, the amounts of inputs, kernel parameters and outputs are independent of *Pop*, *Popc* and *Pks* respectively because of the parallel-reuses. The relation between the input data and *Popc* can be derived using Equation (1), which takes the overlap-reuse into consideration. The total required data is the product of that in all the dimensions.

Therefore, the amount of kernel parameters required by a series of temporal unrollings for each PE can be derived as:

$TP\_K = \prod_{d \in \{B,C,H,W\}} (TP\_Pg_d \times TP\_Pop_d \times TP\_Pks_d)$ (7),

where *TP* means unrolling in the temporal list. When the required amount of kernel parameters exceeds the capacity of KLS (e.g., the last loop that *klst* points to in Figure 9), a data movement occurs to load new data to KLS. Therefore, the number of KLS data movements can be derived as:

$\#KM = \prod_{d \in \{B,C,H,W\}} \prod_{p \in \{ks,opc,op,g\}} out\_klst\_TP\_Pp_d$ (8),

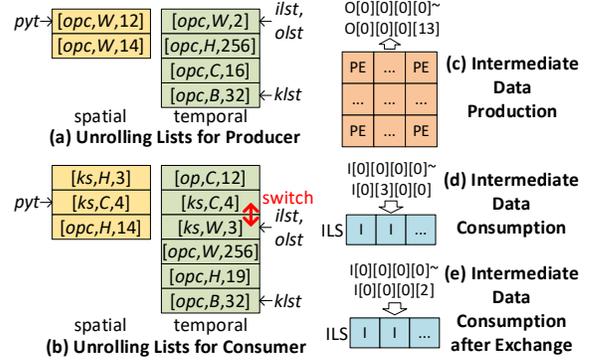

Figure 10. An Example of Unrolling Loop Exchange

where *out_klst_TP* refers to loops outside the *klst*. Similar to Equation (7), the total kernel parameters required by all the working PEs for each cycle is:

$SP\_K = \prod_{d \in \{B,C,H,W\}} (SP\_Pg_d \times SP\_Pop_d \times SP\_Pks_d)$ (9).

Based on Equations (7) to (9), the data movement of KLS is:

$kls\_movement = \#KM \times SP\_K \times in\_klst\_TP\_K$ (10).

The data movement of inputs and outputs and the lower-level memory (e.g., global buffer, off-chip DRAM) can be derived similarly.

### 4.3 Extending to GCONV Chain Acceleration

Besides the algorithm to map a single GCONV operation to a given accelerator, our system also includes two chain optimizations to overcome the challenges to efficiently accelerate the entire GCONV Chain.

**Consistent mapping:** The sharing of global buffer requires the consumer to load the intermediate data in the format stored by the producer. For example, in Eyeriss, outputs unrolled in *px* ($O_1$ and $O_2$ in Figure 8(b)) are generated in parallel and can be collected at the same time while the inputs unrolled temporally ($I_1$ and $I_2$ in Figure 8(a)) can be loaded into the local scratchpads in parallel through the data bus. Therefore, the inner *opc/op/g* loops in *px* unrolling of the producer determine the storage format of intermediate data while the inner *ks/opc/g* loops of the consumer's temporal unrolling determine the optimal loading format. An inconsistent mapping example is illustrated in Figure 10. Based on the mapping of the producer (e.g., DenseNet ReLU1 GCONV1) in Figure 10(a), the buffering format for the intermediate data is shown in Figure 10(c). However, the mapping of the consumer (e.g., DenseNet Convolution2 GCONV1) in Figure 10(b) requires loading the inputs in the format in Figure 10(d), which is not consistent to that in (c).

In GCONV Chain, the intermediate data format inconsistency can be simply solved by loop exchange. For instance, in Figure 10(b), if the unrollings [*ks*, *C*, 4] and [*ks*, *W*, 3] are exchanged, the inputs of the consumer can then be loaded in the format in Figure 10(e). With the original unrolling, only one input is loaded into ILS per cycle. After the exchange, at least three inputs (determined by the unrolling factor and the width of the buffer) can be loaded in parallel. In practice, we also consider exchange of temporal and spatial unrollings of the same parameter as well as unrollings with different parameters in the same unrolling dimension. Additionally, if there are no appropriate ones in the consumer, we check exchange opportunities in the producer. Note that the unrolling loop exchange does not affect the performance or data movement based on Equations (6) and (10) but significantly reduces the loading time for the consumer. In our experiments, this reduces the data loading latency by up to 3.9x compared to the baseline.

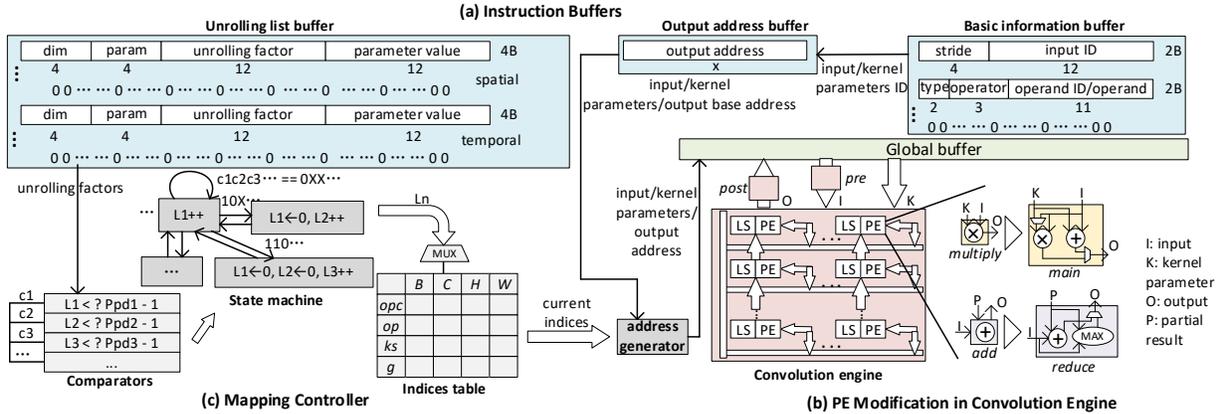

Figure 11. An Overview of GCONV Chain Implementation

**Operation fusion:** Operation fusion is commonly adopted to reduce the movement of intermediate data and to fully exploit the memory bandwidth [39][40]. In GCONV Chain, we also notice an imbalance among the operations in terms of the data/computation ratio. This results in low performance of certain GCONVs with a bottleneck in data loading. Therefore, we apply operation fusion by fusing the GCONVs with no *reduce* operator into the *pre*, *post* or *main* operators of their consumer or producer. For example, GCONV FP2 in Table 2 can be processed as the *post* of FP1 or *pre* of FP3 and FP4. Since the outputs only need to be processed once, fusing to the *post* operator is preferred. After fusion, the *pre* and *post* operators may have more than one parameter and the parameters can be reused in different dimensions. Therefore, to minimize the parameter loading overhead, the consistent mapping also takes the unrolling consistency with the *pre*/*post* operator into consideration.

The operation fusion technique reduces the length of GCONV Chain by up to 30%. It also reduces the total input movement cost by up to 63%. However, due to the *pre*/*post* parameter loading, the kernel parameter movement of the global buffer has increased. On average, operation fusion improves the performance by 1.1x and decreases the data movement energy by 1.3x.

### 4.4 Generalizing to Other Accelerators

Although the GCONV Chain acceleration method proposed is exemplified by Eyeriss, it easily generalizes to other CNN accelerators. The performance analysis in Section 4.2 intrinsically applies to all the cases. Therefore, the exact mapping algorithm of a random accelerator only relies on its specific structure for unrolling. In our exploration, all the accelerators manifest both the spatial and temporal unrolling dimensions. The difference lies in the number and functions of the spatial dimensions as well as the capacity and hierarchy of the memory. Among the evaluated accelerators in Table 4, [17][26] possess two spatial dimensions, one with input parallel-reuse and the other with *reduce* but no overlap-reuse; [25] has two spatial dimensions with one for overlap-reuse; and the only spatial dimension in the subsystem of [6] can exploit *reduce* and overlap-reuse at the same time. In terms of memory, most accelerators adopt two-level on-chip storage. For those accelerators with no local scratchpads (e.g. [21]), the sizes of the local scratchpads can be set to 1. In some accelerators, only a certain data type has a local memory (e.g., the input pool in [6]).

Despite the variance in the structure of the accelerators, the underlying mapping philosophy, i.e., to first occupy the spatial dimensions and the special functions to maximize the performance and data reuse, remains the same. In the evaluation, we follow the mapping strategies provided in the original works of the baselines, which just slightly changes the priority of the parameters in Lines 7 to 22 of Algorithm 1 in section 4.1. The two chain optimizations in Section 4.3 do not rely on a certain accelerator either. For a given structure, we just need to recognize the output and input format determining dimensions to guarantee that the inner loops in these two dimensions are consistent.

## 5. GCONV Chain Implementation

To apply GCONV Chain to an existing accelerator, necessary supports are inserted into the computation stack.

First, we implement a compiler that automatically transforms a neural network into a GCONV Chain and then performs optimizations and mapping based on the given accelerator structure. Our compiler is implemented in Python and all the networks and hyperparameters are extracted from Caffe [41] through the Pycaffe interface. For all the CNNs and accelerators, it takes an average of 0.024 seconds to transform and auto-map one layer. This generates a list of GCONV instructions, which are executed by the GCONV-augmented accelerator shown in Figure 11.

Table 4. GCONV Chain Implementation Configurations

| Category | Accelerator | Configuration | PEs | Local Storage | Global Buffer | Bandwidth |
|---|---|---|---|---|---|---|
| TIP | TPU [21] | 64 rows, 64 columns | 4096 | ILS: 1 per PE, OLS:1 per PE, KLS:1 per PE | I & O: 1.5MB K: 0.25MB | I: 64, O: 64, K: 11 |
| LIP | DNNWeaver (DNNW) [25] | 14 PUs, 74 PEs per PU | 1036 | ILS: 1 per PE, OLS: 1 per PE, KLS: 1 per PU | I & O & K: 8.5kB per PE, K: 8.5kB per PU | I & O & K: 1 for 2 PEs, K: 1 per PU |
| CIP | Eyeriss (ER) [5] | 12 rows, 14 columns | 168 | ILS: 12 per PE, OLS: 24 per PE, KLS: 224 per PE | I & O: 100kB, K: 8kB | I: 1, O: 4, K: 4 |
| CIP | EagerPruning (EP) [6] | 4 subsystems, 512 PEs per subsystem | 2048 | ILS: 64 per subsystem, OLS: 1 per PE, KLS: 1 per PE | I: 1.5MB, O: 1.5MB, K: 1.5MB | I: 32 per subsystem, O: 32 per subsystem, K: 32 per subsystem |
| CIP | NLR [7] | Tm = 64, Tn = 7 | 448 | ILS: 1 per Tn, OLS: 1 per Tm, KLS: 1 per PE | I & K: 1.5MB, O: 0.75MB | I & K: 7, O: 64 |

Figure 11(a) shows the instructions of a GCONV operation. There are three instruction buffers in the system. The basic information buffer stores the stride, operators, input and kernel parameter producer IDs. Considering that some GCONVs do not have *pre*, *main*, *reduce* or *post* operators, the first field of the operator instruction is utilized to indicate the operator type. An all-zero entry delimits the basic information of the GCONVs. For the unrolling list buffer, the first three fields are the unrolling dimension, parameter and unrolling factor respectively, as in Figure 9, while the last field indicates the argument of the parameter. If the parameter is unrolled more than once, the argument is the sum of all the entries that unroll the same parameter. The unrolling lists in different unrolling dimensions for the GCONVs are also delimited by an all-zero entry. The last instruction buffer stores the address of the output generated by each GCONV, the width of which is determined by the size of the data buffer.

The accelerator is equipped with a set of registers to buffer the stride, parameters, operators and unrolling lists. In the setup stage of each GCONV, one instruction entry is read from the basic information and unrolling list buffers in each cycle. The decoder translates the instructions dictated by a state machine. During the process, the last entry (e.g. *pyt*, *ilst*) of each dimension and the arguments of the parameters are generated while decoding the unrolling lists. The addresses of input and operands for the operations are derived by indexing the IDs in the output address buffer and the output address is allocated in run-time based on the current data buffer occupation and the size of the output. To eliminate the possible delay, instruction loading and decoding are overlapped with the processing of the previous GCONV.

As mentioned in Section 3.1, GCONV does not change the inherent connections in the convolution engine. The only modification is to replace the original *multiply* and *add* functions with comprehensive *main* and *reduce* functions, as shown in Figure 11(b). During GCONV processing, the loop iterations are carried out by a state machine, as shown in Figure 11(c). Since the unrolling lists are not fixed, it is impossible to use a predefined state machine. Instead, the transition conditions are set as the results of comparison between the unrolling factors and the counters. A 16:1 MUX is adopted to increase the index of the corresponding parameter. The address generator generates the offset of the data based on the index and data storage format. Then the data loading module loads the data and feeds them to the PEs through the data bus.

## 6. Evaluation

### 6.1 Baselines and Benchmarks

For a comprehensive evaluation, we include all three types of CNN accelerators discussed in Section 2.3 which are listed in Table 4. Most of the baseline cases adopt the dataflow and the PE and memory configurations as in the original works [21][22][26]. Note that we perform dense computation on EP to focus on the hardware acceleration. TPU is proposed as a datacenter-level design, so we scale down its basic block in [21] by 4×4 to match the other accelerators. DNNW generates a customized accelerator for each CNN. We adopt the configuration of AlexNet, the only benchmark we share, on the moderate FPGA Altera Stratix V SGSD5 in [25]. For GCONV Chain implementation, all the layers are converted into GCONV operations and auto-mapped to the convolution engine (or matrix functional units in TIPs) of the accelerators. Considering that the LIP baseline allocates computation resources to some other dedicated functional units, which will be idle in GCONV Chain processing, the convolution engine of the GCONV Chain implementation is scaled up so that they have the same number of PEs and total available bandwidth as the baselines. Columns 3-7 of Table 4 summarize the GCONV Chain configurations of the accelerators.

For the benchmarks, we evaluate the seven CNNs in Table 1(a). Note that ZFFR, CapNN and C3D are not evaluated on baseline DNNW and C3D is not evaluated on all the CIP baselines since on-chip acceleration of the functional layers in these networks is unclear in the original papers. In the experiment, we focus on the training of CNNs, which includes the computation in inference and provides more insights.

### 6.2 Methodologies

To demonstrate the three benefits brought by GCONV Chain discussed in Section 1, we study the speedup, overhead, energy efficiency and whole-life cost of GCONV Chain. Specifically, Section 6.3 evaluates the speedup of GCONV Chain over baselines to show that it can apply to any accelerator. Sections 6.5 and 6.6 compare the energy efficiency of

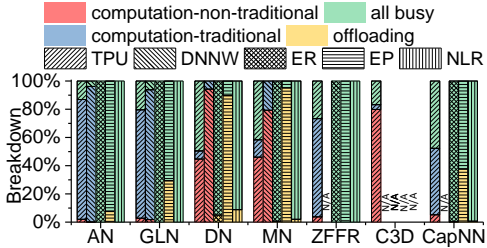
Figure 12. Baseline Latency Breakdown

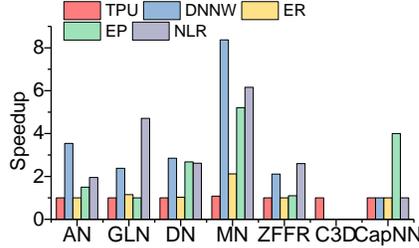
Figure 13. Convolution Layers Speedup

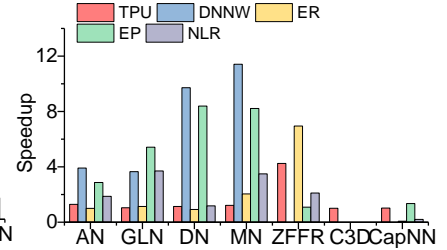
Figure 14. End-to-end Speedup

GCONV Chain-armed CIPs with TIPs and LIPs to show its potential in low-cost CNN acceleration.

We develop a simulator to evaluate the performance and data movement based on the model proposed in Section 4.2, which is validated on a cycle-accurate basis. To get the area and energy estimation, we prototype the accelerators and synthesize the RTL using Synopsys Design Compiler and simulate the memory with CACTI [42].

All the accelerators run at 700MHz. For the baselines, the operations of all the layers are processed on-chip in TIPs and LIPs while in CIPs, only the traditional layers mentioned in Section 2.1 are processed on-chip while the other layers are offloaded to an ARM A53 CPU through PCIe 4.0. For computations allocated to different functional units or the host, they are processed in a pipeline to keep the maximal resources busy.

### 6.3 Speedup
The end-to-end speedup brought by GCONV Chain comes from two aspects: (1) it eliminates the inefficiencies of the baselines in terms of processing the non-traditional layers; (2) for the most computation-intensive convolution layers, GCONV Chain can still improve the performance thanks to its flexible mapping.

To implicate the inefficiencies of the baselines, Figure 12 first shows the latency breakdown of them. Among the baseline accelerators, TPU and DNNW suffer from pipeline bubbles with considerable time only running either the traditional or non-traditional layers (computation-traditional or computation-non-traditional in Figure 12). The runtime that all the components are busy (all-busy) only accounts for 31% and 2% in TPU and DNNW respectively. The utilization is higher in TPU because it accelerates fine-grained tensor operations while the instructions in DNNW are more complex. EP suffers the most from the offloading (43% of runtime on average) because it has the highest on-chip performance. While ER and NLR can overlap the offloading by computation to some extent, the offloading power is not negligible as will be shown in Figure 18. In terms of each CNN, the offloading latency is more severe in recent CNNs with more non-traditional layers (e.g., DN, MN). However, CNNs with non-traditional layers highly concatenated (e.g., ZFFR, C3D, CapNN) suffer less from offloading.

Figure 13 shows the speedup of the convolution layers to demonstrate the effectiveness of GCONV mapping. In all the cases, the performance of GCONV Chain is no worse than the baselines. In MN, where the feature maps unrolling in the baselines is useless for depthwise convolution, the speedup is salient. GCONV Chain also significantly speeds up the convolution layers in baseline NLR, which only unrolls the input and output feature maps. The speedup over baseline TPU and ER are low because they explore flexible unrolling strategies. EP is similarly flexible as ER but the huge PE array makes the baseline mapping less effective. Fortunately, GCONV Chain manages to improve its performance.

When it comes to the end-to-end CNN acceleration including all the traditional and non-traditional layers, Figure 14 shows the speedup of GCONV Chain to the baselines. The results show that GCONV Chain speeds up the baselines by up to 8.2x and an average of 3.4x among all the accelerators. The speedup of DN and MN on DNNW and EP are high because their baselines suffer the most from the pipeline bubbles and offloading. The speedup of CapNN on ER and NLR is low because their on-chip computing power cannot compare to that of A53.

### 6.4 Overhead
We aim to compare the total cost of GCONV Chain-armed CIPs (GC-CIPs) with LIPs and TIPs, so this section focuses on the overheads brought by GCONV Chain to CIPs. Figure 15 compares the average code length of GC-CIPs with LIPs and TIPs. On average, GC-CIPs instructions are 5.8x longer than LIPs because LIPs have only one instruction for each layer. TIPs only process basic matrix or vector algorithms, so control operations are needed when the computation cannot be mapped to only one matrix/vector operation. In addition, they require load instructions while LIPs and GC-CIPs load data implicitly. Therefore, their code density is the worst (2.6x worse than GC-CIPs).

Figures 16 and 17 list the overhead of GCONV Chain in the area and the average power breakdown of Eyeriss. The storage overhead refers to the storage for the instruction buffers in Figure 11(a) and the compute overhead corresponds to the PE modification in Figure 11(b). The control overhead includes all the required signals in Figures 11(a)(b) and the controller in Figure 11(c). In total, GCONV Chain brings 20% area and 19% power consumption overhead. This is acceptable considering the speedup and reduction in data movement.

### 6.5 Energy Efficiency
In CNN accelerators, it is widely recognized that the data movement dominates the energy efficiency [24]. Figure 18 presents. The measurement includes the on-chip global buffer movements and offloading and reloading related power normalized to the baseline of TPU. The off-chip data movement is not considered because GCONV Chain does not substantially affect the off-chip data access in our experiments. As shown, although the on-chip data movement reduction brought by GCONV Chain is not significant, it eliminates the costly offloading and reloading of non-traditional layers in CIPs. Compared with TIPs, GC-CIPs that explore more data reuses have the lowest data movement (16% and 22% in ER and EP). Note that NLR does not have low on-chip data movement because it does not exploit any overlap-reuse.

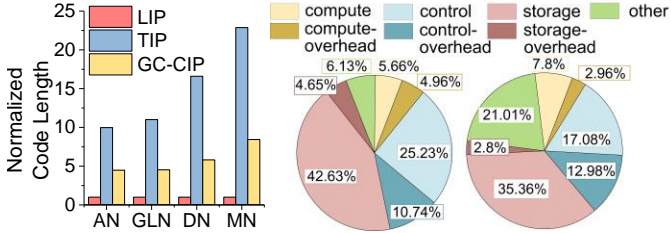
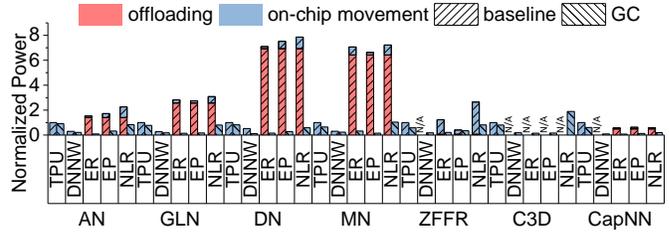

Figure 15. Code Length Comparison

Figure 16. Area Breakdown

Figure 17. Power Breakdown

Figure 18. On-chip and Offloading Data Movement Power

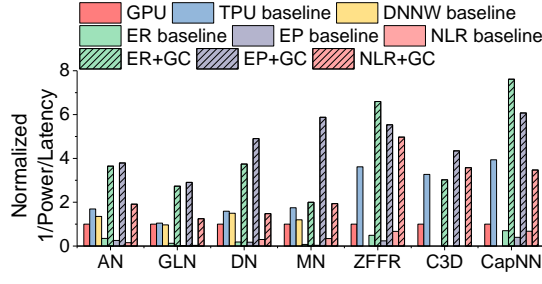
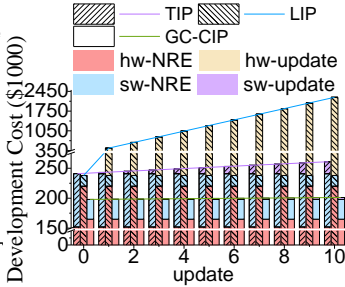
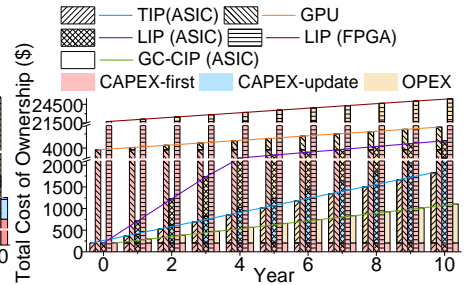

Figure 19. Energy Efficiency (Iso-Power Performance)

Figure 20. Development Cost

Figure 21. Total Cost of Ownership

Figure 19 further shows the normalized overall energy efficiencies of the GC-CIPs, TIP, LIP and a state-of-the-art GPU, i.e., NVIDIA Tesla V100. Equipped with GCONV Chain, the CIP accelerators with overlap-reuse (i.e., ER and EP) overcome the inefficiency in the baselines (37.6x on average) and show a promising edge over TIP (up to 3.4x, 2.1x on average), LIP (up to 4.9x, 3.0x on average) and GPU (up to 7.6x, 4.5x on average).

### 6.6 Whole-life Cost

Last but not the least, we compare the whole-life costs of TIP, LIP and GC-CIP. Figure 20 shows the development costs as a sum of hardware/software non-recurring expenses (NRE) and update costs. Based on the complexity level of the accelerator implementation, the hardware NRE of TIPs, GC-CIPs and LIPs are quoted as 152K, 165K and 220K USDs [43]. Then in each update, LIPs require 200K USDs on the new hardware design. The software NRE and update costs are calculated using the latest software engineer salary [44][45] and the lines of code in our prototype compiler. Although GC-CIPs consume more in the hardware than TIPs, the software development is cheaper due to the code generation complexity of TIPs. This gap widens as more updates are performed and 60K additional USDs are consumed for development of TIPs than GC-CIPs after ten updates.

Users who invest in the accelerators need to pay the capital expenses (CAPEX) for the device purchase and annual update and the utility as operating expenses (OPEX). Figure 21 shows the total costs of ownership for the ASIC version of the three types of accelerators as well as FPGA LIPs and GPUs, which are popular choices for CNN acceleration. The CAPEX of FPGA and the ASICs are scaled to meet the performance of GPU and the operating utility is calculated assuming the devices are always working at the average utility rate in US [46]. As observed, the GPU, FPGA and ASIC LIPs with high CAPEX [43][47][48] are not the best choices for pure CNN acceleration. Thanks to the high energy efficiency of convolution customized dataflows, GC-CIPs win the most whole-life efficient CNN accelerators by costing 45% less than TIPs after just three years and 65% less after ten years.

## 7. Related Work

Besides the accelerators discussed in Section 2.3, there are several works trying to efficiently process the non-traditional computation in CNNs. [8] infuses the adder tree with pooling functions and [11] adds local response normalization layer support but the other layers are still left behind. [40] introduces a method to break down the batch normalization into two parts and fuse the computation to the next and last layers in the CNNs, which is orthogonal to our work and can be adopted when optimizing the GCONV Chain. [49] introduces batch size as a tunable parameter in CNN accelerators to make up for the lack of parallelism in traditional convolution acceleration but it does not systematically accelerate all the non-traditional layers. [39] is proposed to assist mapping neural networks defined in any framework to any hardware. However, it currently only supports matrix/vector operations and commercial general-purpose processors. [50][51] propose models to explore the mapping design space of convolution but they only focus on the convolution layers.

## 8. Conclusions

This work aims to address an overlooked challenge: *the efficient and cost-effective acceleration of the diverse end-to-end CNN computation*. To exploit the reuse opportunities in CNNs and to avoid resource underutilization and costly upgrade brought by allocating dedicated hardware to the non-traditional layers, we propose to build a general convolution model and generalizing the diverse computations in CNNs into GCONV Chain. By generalization, the end-to-end GCONV Chain can be efficiently processed by existing accelerators customized for convolution. Our evaluation shows that GCONV Chain manifests great potential in processing the emerging CNNs on existing customized accelerators with *high performance, energy efficiency and low whole-life cost*.